\documentclass[aps, prl, twocolumn, showpacs, floatfix]{revtex4}
\usepackage{graphicx}

\begin{document}

\title{Half Metallicity in Hybrid BCN Nanoribbons}

\author{Er-Jun Kan$^1$}
\author{Xiaojun Wu$^2$}
\author{Zhenyu Li$^1$}
\author{X. C. Zeng$^{2*}$}
\author{Jinlong Yang$^{1}$}
\thanks{Corresponding author. E-mail: jlyang@ustc.edu.cn; xczeng@phase2.unl.edu}
\author{J. G. Hou$^{1}$}

\affiliation{$^1$Hefei National Laboratory for Physical Sciences at
     Microscale,  University of Science and Technology of
     China, Hefei,  Anhui 230026, China \\
     $^2$Department of Chemistry and Nebraska Center for
     Materials and Nanoscience, University of Nebraska-Lincoln, Lincoln, Nebraska 68588, USA}

\date{\today}

\begin{abstract}
We report a first-principles electronic-structure calculation on C
and BN hybrid zigzag nanoribbons. We find that half-metallicity can arise in the
hybrid nanoribbons even though stand-alone C or BN nanoribbon
possesses a finite band gap. This unexpected half-metallicity in the
hybrid nanostructures stems from a
competition between the charge and spin polarizations, as well as from the
$\pi$ orbital hybridization between C and BN. Our results point out a
possibility of making spintronic devices solely based on
nanoribbons and a new way of designing metal-free half metals.
\end{abstract}


\maketitle

The discovery of low-dimensional carbon allotropes such as
fullerenes \cite{c60} and nanotubes \cite{cnt} has stimulated
intensive research on metal-free magnetism, owing to their small
spin-orbit coupling and long spin scattering length. Since the
first report of room-temperature weak ferromagnetism in polymerized
C$_{60}$ \cite{c60m}, magnetism in pure carbon materials has been
demonstrated by several experimental groups \cite{cm1,cm2,cm3}.
Carbon defects or adatoms are believed to be the origin for the
unexpected magnetism  \cite{tcm1,tcm2,tcm3}.

A question arises that
can low-dimensional carbon nanostructures exhibit half metallicity?
Half metals are ideally for
spintronic applications because they have
 one metallic spin channel and one semiconducting or insulating spin
channel. Techniques for tuning electronic properties of
graphene nanoribbon (GNR)
have been advanced \cite{gnr1,gnr2,gnr3}, for example, by terminating
a single graphite layer in one direction. Recently,
Son {\it et al.} \cite{hf1} predicted, based on density functional theory (DFT)
calculation, that zigzag edge GNRs (ZGNRs) can be converted to half metal when an external
transverse electric field is applied onto the graphene layer. Our recent study using
hybrid DFT confirmed their prediction\cite{hf2}. However, we found that
 a very strong field is required to achieve
half metallicity, a limitation for wide application.

In the absence of external field, ZGNRs are semiconductor with two localized
electronic edge states \cite{z1,z2,z3,z4,z5}. These two
ferromagnetically ordered edge states are antiferromagnetically
coupled. In the presence of an external field, an electrostatic potential
difference is generated between the two ZGNR edges, which causes band crossing
and charge transfer between the two edge states. As a result, the
ZGNRs are converted to half metal from semiconductor. Since the critical
step to make half-metallic ZGNR is to break the symmetry
of the two edge states, an alternative approach is to use
chemical decoration. For nanoelectronic applications, the
chemical approach has many advantages over the strong-field
approach.

In this Letter, we present two nanostructure designs to achieve
half-metallicity for carbon nanoribbons without using metal
dopants. The first design is by implanting a BN row into a ZGNR.
The principle of this materials design is based on the following
observation: both the highest valance band (VB) and the lowest
conduction band (CB) of ZGNR are mainly contributed by the two
edge states which have opposite spin but are degenerate in energy.
In contrast, for the zigzag BN nanoribbon, the highest VB is
originated from the $\pi$ orbitals of the N edge atoms, while the
lowest CB is mainly localized at the B edge atoms. These results
suggest that the symmetry of the ZGNR band structure near the
Fermi energy can be broken by introducing a segment of BN
nanoribbon.

\begin{figure}[!hbp]
  \includegraphics[width=8cm]{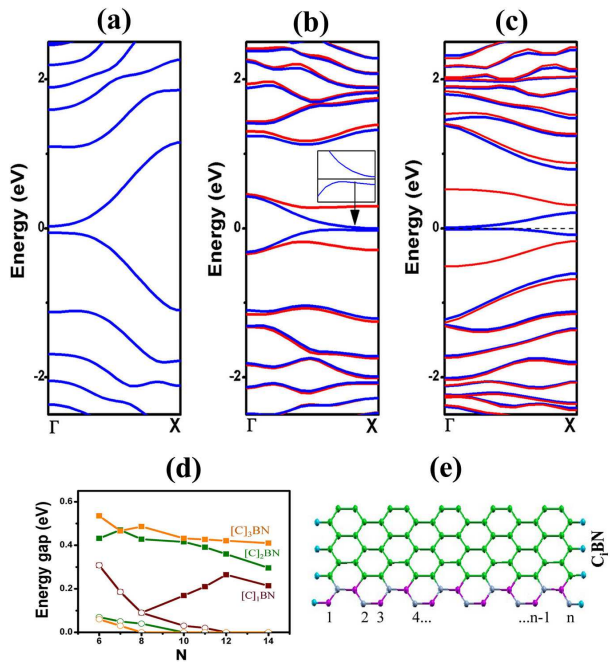}
  \caption{ Band structures for (a) C$_{1}$BN, (b)
  C$_{2}$BN, (c) C$_{3}$BN with 8 zigzag chains. The red (blue) lines
  represent the spin-up (spin down) channel.
  (d) The spin-resolved band gaps of $n$-C$_{i}$BN nanoribbons.
  The filled square and hollow circle denotes the spin-up and spin-down channel, respectively. (e)The atomic configuration of C$_{i}$BN ribbons.
  Green, pink, grey, and sapphire balls denote carbon, boron, nitrogen,
  and hydrogen atoms, respectively.}
  \label{fig1}
\end{figure}

A geometry of the newly designed hybrid C/BN nanoribbon is shown
in Fig~\ref{fig1}e, where the armchair atom rows are normal to the
ribbon direction while
 the zigzag atom chains are in the ribbon direction. At the two edges,
H atoms are added to passivate the
dangling $\sigma$ bonds. The hybrid ribbon structures are denoted as
$n$-C$_{i}$BN, where $i$ refers to the number of armchair C
rows in a unit cell, and $n$ is the width of the nanoribbon in term of the
number of zigzag chains. On one edge, B is the outmost
non-hydrogen atom; this edge is denoted as the B-edge. The adjacent C atoms to the
B-edge are named C$_B$. Similarly, the C atoms adjacent to
the N-edge are named C$_N$.


The DFT calculations were carried out using the VASP package \cite{vasp}.
The projector augmented wave
(PAW) \cite{paw} method in Kresse-Joubert implementation
\cite{K} was used to describe the electron-ion interaction. The
plane wave cut-off energy was set to $400.0$ eV. The convergence
thresholds for energy and force were 10$^{-5}$ eV and 0.01 eV/{\AA},
respectively. Perdew-Wang functional \cite{pw} known as PW91 was
used in the generalized gradient approximation (GGA). In a previous
study of ZGNRs, we showed that the GGA gives
qualitatively the same results as hybrid density functional
\cite{hf2}. Here, the calculated band gaps of the pristine ZGNR and BN
nanoribbon with 8 zigzag atom chains are 0.43 and 4.2 eV
respectively, in good agreement with previous theoretical
calculations \cite{hf1,tc1}.

The calculated electronic structures of 8-C$_i$BN ($i$=1, 2, 3)
are shown in Fig 1. The band structures of 8-C$_1$BN and 8-C$_3$BN
exhibit different patterns compared to 8-C$_2$BN, which can be
understood from the Brillouin-zone folding. When $i$ is even, the
largest gap locates at $\Gamma$ point, and when $i$ is odd the
smallest gap locates at $\Gamma$ point. 8-C$_1$BN has a
nonmagnetic ground state with a small band gap of $\sim$0.1 eV.
Both 8-C$_2$BN and 8-C$_3$BN are spin-polarized, and the latter is
half metal since its valence and conductive bands overlap at the
Fermi level in the spin-down channel.  This half-metallicity is
unexpected, considering that both C and BN zigzag nanoribbons
possess a finite band gap. We further examined nanoribbons with
different widths ($n = 6-14$). As shown in Fig 1d, all C$_i$BN
(i=1, 2, 3) nanoribbons become half metal when $n \ge 12$. The
critical widths are $n = $12, 10, and 8 for $i$=1, 2, and 3,
respectively.

The calculated partial density of states (PDOS) indicates that the
8-C$_1$BN has no spin polarization. Near the Fermi level, the
occupied states are contributed by C$_N$ and N edge atoms, while
the unoccupied states are contributed by C$_B$ and B edge atoms.
This is a charge polarized state with electron transfer from C$_B$
to C$_N$. The charge polarization is clearly seen in Fig. 2a-g;
the partial charge density from 0.05 eV below the
Fermi level to the Fermi level is localized at the N-edge.



In pristine ZGNR, every edge C atom at the edge has only two C
neighbors and thus 1/3 nonbonded extra $\pi$ electron, which gives
rise to the edge states and a high electron density at the Fermi
level. A stablization mechanism is required to have a lower-energy
ground state \cite{z3}. In the charge-polarization mechanism, the
extra electron transfers completely from one edge to the other. In
the spin-polarization mechanism, the C atoms at the two opposing
edges have spin-opposite electrons, which lowers the on-site
Coulomb interaction $U$ between electrons with opposite spin at
the same site. The spin mechanism is more favorable in energy
\cite{z3}. However, the charge-polarized state can be further
stabilized by an external electric field. Hence, a ZGNR can be
changed from having spin polarized state to charge polarized state
under external field. In either the spin or charge polarized
state, a ZGNR is semiconductor. Here, the half-metallicity for
ZGNR is found in the region where the two states compete with each
other\cite{hf2}.

\begin{figure}[!hbp]
  \includegraphics[width=8cm]{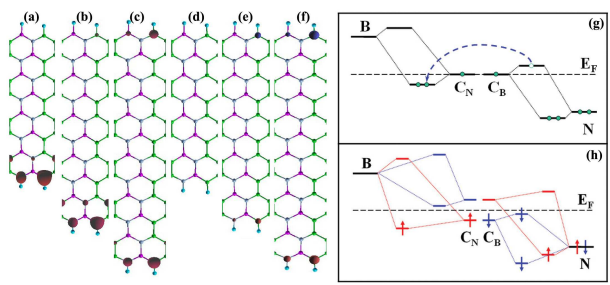}
  \caption{The partial charge density within energy range [E$_{f}$-0.05, E$_{f}$] in eV
  for $n$-C$_{1}$BN with $n$ equals to (a) 8, (b) 10, and (c) 12. Isovalue
  is 0.02 e/{\AA}$^{3}$.
  Spin density of $n$-C$_{1}$BN with $n$ equals
  to (d) 8, (e) 10, and (f) 12. (g) and (h) are schematic energy diagram for C$_i$BN. (g)
  Charge polarization caused by $\pi$ electron hybridization between C and BN.
  The blue arrow indicates the direction of electron transfer.
  (h) Competition between spin and charge polarization.
  Spin degenerated orbitals are represented by long level lines, and spin
  orbitals are denoted by short level lines.  Red and blue represent the spin
  up and spin down channel, respectively.
  }
  \label{fig2}
\end{figure}

The charge-ordered ground state of 8-C$_1$BN is stablized by the
mixing of $\pi$ orbitals of C and BN. In C$_1$BN, each C$_B$ atom
is covalently bonded to a N atom, while each C$_N$ atom is
connected to a B atom. As shown in Fig. \ref{fig2}g, the occupied
$\pi$ orbital of N is much lower in energy than the unoccupied
$\pi$ orbital of B. The $\pi$ orbitals of C locate in the middle.
The interaction between the $\pi$ orbitals of C$_B$ and N results
in a higher antibonding orbital. However, the electron tends to
occupy the lower bonding orbital between C$_N$ and B. Such an
orbital interaction leads to electron transfer from C$_B$ to
C$_N$.

As the nanoribbon width ($n$) increases, the Coulomb interaction
between the two edge states decreases, and thus the stabilization energy of the
charge-polarized state also decreases. On the other hand, the
on-site Coulomb interaction is a local property, and the
spin-polarization energy is independent of the ribbon width
\cite{z3}. Hence, the spin polarization becomes
relatively more competative as the nanoribbon width increases.

As shown in Fig~\ref{fig2}b, when the ribbon width is not wide
enough (e.g., $n=10$), the ribbon still has an occupied N-edge and
nearly empty B-edge state. However, the spin degeneracy is already
broken. The spin-up and spin-down electrons tend to occupy one of
the two subsites of the honeycomb lattice, which leads to the
energy of spin-down edge state higher than of the spin-up one.
When $n$=12, the occupied spin-down state at the N-edge and the
unoccupied spin-down state at the B-edge cross over in energy. The
B-edge is partly occupied, which weakens the charge polarization.
More importantly, in this case, the spin-down channel becomes
metallic. Therefore, the competition between the spin and charge
polarization leads to half-metallicty in this hybrid-ribbon
structure. As the ribbon width increases, a decrease of charge
polarization (Fig. \ref{fig2}a-c) and an increase of spin
polarization (Fig. \ref{fig2}d-f) can be clearly seen.

For C$_i$BN with $i > 1$, the $\pi$ orbital hybridization only
occurs between BN and its nearest neighboring C atom row. In the
$i$=$\infty$ limit, the electronic structure of C$_i$BN will
converge to that of ZGNR. For different $i$, we analyzed the
competition between charge and spin polarization and the
half-metallicity based on the spin-polarized ZGNR ground state
(see Fig. \ref{fig2}h).
 In ZGNR, C atoms at the two edges are
degenerate in energy but with opposite spin. When the BN row is
introduced, the energy of the bonding $\pi$ orbital between C$_N$
and B is lower than that of the edge state in ZGNR. This
bonding state is occupied by spin-up electrons. On the other
hand, for C$_B$ and N, both the bonding and anti-bonding states are
occupied by the spin-down electrons. Hence, in C$_i$BN
without charge polarization, the valence band maximum (VBM) stems from
the anti-bonding states between N and C$_B$ in the
spin-down channel, while the conduct band minimum (CBM) is composed
of the bonding states between B and C$_N$ in the same spin
channel. With the increase of the ribbon width, the ZGNR band gap is
reduced, and the VBM and CBM are closer in location. Once the VB and CB overlaps
at a specific point ($i$=3 and $n$=8), spin-down
electron at the B-edge will transfer to N-edge. Consequently, at both edges,
the spin-down channel is partially occupied, and the partial charge
polarized nanoribbon becomes half metal.

\begin{figure}
 \includegraphics[width=8cm]{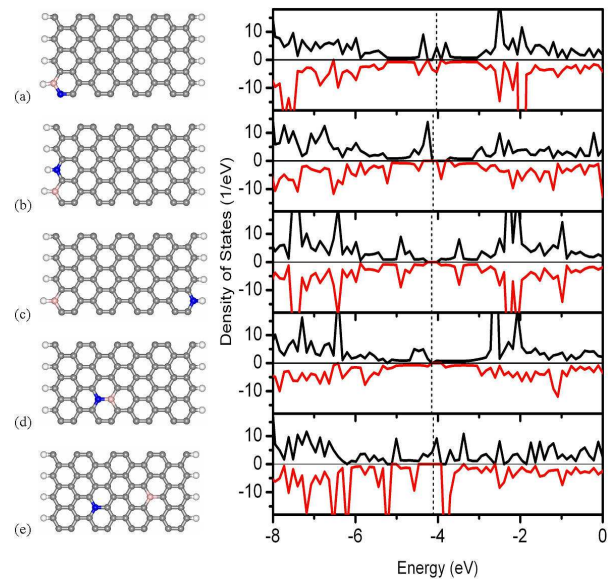}
 \caption{ (a) $\sim$ (e) Configurations and density of states of a pair of B and N atoms
 substituted ZGNR. Fermi level is plotted with dotted line. Grey, blue,
 khaki and white balls denote carbon, nitrogen, boron, and hydrogen atoms, respectively.}
 \label{fig3}
\end{figure}

Note that implantation of a BN row in carbon nanotube has
been realized in the laboratory \cite{Stephan9483, Enouz0756}, but
not yet in ZGNR. Our first design for the hybrid C/BN
nanoribbon is a highly ordered nanostructure. It may be challenging to
realize this nanostructure in the laboratory but it can serve
 as a proof of principles for achieving metal-free half
metal. For the purpose of practical (mass) production of the
ZGNR-based half metal, we propose an alternative material design
as illustrated in Fig. \ref{fig3} (left panels). Here, a pair of B
and N atoms are doped in place of two C atoms in each supercell of
ZGNR (dilute substitution). The location of the substitution is
arbitrary. In Fig. \ref{fig3} (right pannels), we show the
calcualted DOS for five typical configurations of the BN
substitution, three at the edge and two in the interior of the
ZGNR. It can be seen that with the exception of the symmetric
substitution at the opposing edge [Fig. \ref{fig3}(c)], four out
of five configurations exhibit spin polarization also. Some hybrid
nanoribbons can be viewed as spin-selective semiconductor [e.g. in
the case of Fig. \ref{fig3}(b)].

\begin{figure}
 \includegraphics[width=8cm]{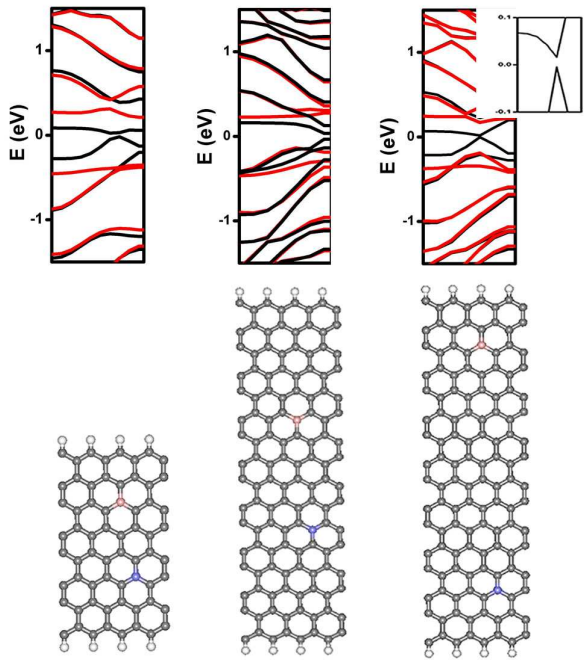}
 \caption{ Band structures of randomly doped ZGNR
 with different width.  Red for the
 spin-up channel, and black for the spin-down channel.Grey, blue, khaki and
 white balls denote carbon, nitrogen, boron, and hydrogen atoms, respectively.}
 \label{fig4}
\end{figure}

The most interesting case is the configuration
 shown in Fig. \ref{fig3}(e), where both B and N atoms are located in the interior
of the ZGNR but not as a nearest neighbor. Appreciable DOS can be
seen near the Fermi level in the spin-up channel whereas a finite
band gap can be seen in the spin-down channel. The calculated band
structure (left panel in Fig. \ref{fig4}) shows that there is a
very narrow band gap in the spin-up channel (about a few tens of meV).
Hence, this heterogeneous nanoribbon exibits half-metal or
half semi-metal like
behavior. This result suggests that half metallicity can be
achieved by doping B and N atoms within the interior region of
the ZGNR. The doping can be random as long as it is dilute to
assure certain separation between the B and N atoms. To confirm
this design, in Fig. \ref{fig4} (middle and right panels), we show
the calculated band structures for two additional doping
configruations in a wider ZGNR. Evidently,
 the band gap in one spin channel is narrower.
This result suggests that electrons in this spin channel behave
like metal (or semi-metal), especially when the the B and N atom
is located near the two opposing edges, respectively (right panel
in Fig. \ref{fig4}).

In conclusion, we have studied spin-dependent electronic properties of
 heterogeneous C/BN nanoribbons using density
functional theory. We find that a ZGNR can be converted to half metal (or half
semi-metal) either by inbedding  an
ordered BN row in ZGNR or by substituting B and N atoms in the interior of
ZGNR. In both cases, the $\pi$ orbital hybridization between C and B (or
N) breaks the symmetry of the two edge states. Finally,
we note that in both material designs no metal dopants are involved. This
metal-free half-metallicity points to a new venue
for making spintronic/electronic devices at low-cost.

\begin{acknowledgements} This work is partially supported by the
National Natural Science Foundation of China (50121202,
20533030,20628304), by National Key Basic Research Program under
Grant No.2006CB922004, by the USTC-HP HPC project, and by the
SCCAS and Shanghai Supercomputer Center. The UNL group is also
supported by  US DOE (DE-FG02-04ER46164), NSF,
and the Nebraska Research Initiative.

\end{acknowledgements}

\end{document}